\documentclass{nature-preprint}
\usepackage{tcolorbox,color}
\usepackage{colortbl}
\usepackage{ulem}
\usepackage{amsmath}
\usepackage{amssymb}
\usepackage{graphicx}
\usepackage{upgreek}

\usepackage{caption,wrapfig}
\usepackage{hyperref}

\title{Optical reconfiguration and polarization control in semi-continuous gold films close to the percolation threshold}

\author{Christian~Frydendahl$^{1,2}$, Taavi~Rep{\"a}n$^{1}$, Mathias~Geisler$^{1,2}$, Sergey~M.~Novikov$^3$, Jonas~Beermann$^3$, Andrei~Lavrinenko$^1$, Sanshui~Xiao$^{1,2}$, Sergey~I.~Bozhevolnyi$^3$, N.~Asger~Mortensen$^{1,2,3}$ \& Nicolas Stenger$^{1,2}$}

\begin{document}
\maketitle

\begin{affiliations}
\item Department of Photonics Engineering, Technical University of Denmark, \O rsteds Plads 343, DK-2800~Kongens~Lyngby, Denmark\\
\item Center for Nanostructured Graphene, Technical University of Denmark, \O rsteds Plads 343, DK-2800 Kongens Lyngby, Denmark\\
\item Centre for Nano Optics, University of Southern Denmark, Campusvej 55, DK-5230~Odense~M, Denmark
\end{affiliations}

\begin{abstract}

Controlling and confining light by exciting plasmons in resonant metallic nanostructures is an essential aspect of many new emerging optical technologies. Here we explore the possibility of controllably reconfiguring the intrinsic optical properties of semi-continuous gold films, by inducing permanent morphological changes with a femtosecond (fs)-pulsed laser above a critical power. Optical transmission spectroscopy measurements show a correlation between the spectra of the morphologically modified films and the wavelength, polarization, and the intensity of the laser used for alteration. In order to understand the modifications induced by the laser writing, we explore the  near-field properties of these films with electron energy-loss spectroscopy (EELS). A comparison between our experimental data and full-wave simulations on the exact film morphologies hints toward a restructuring of the intrinsic plasmonic eigenmodes of the metallic film by photothermal effects. We explain these optical changes with a simple model and demonstrate experimentally that laser writing can be used to controllably modify the optical properties of these semi-continuous films. These metal films offer an easy-to-fabricate and scalable platform for technological applications such as molecular sensing and ultra-dense data storage. 

\end{abstract}

\section*{Introduction}
The ability of metallic nanostructures to localize and enhance optical fields down to the nanoscale via collective electron excitations (plasmons) has been the subject of intense study in the recent decades\cite{Schuller:2010}. Many promising applications arise from the careful engineering of nanostructures to tune their optical properties. This allows sub-diffraction light focusing for spectroscopy and sensing applications like surface-enhanced Raman scattering (SERS)\cite{Jeon:2016}, enhancing light-matter interaction with new 2D materials\cite{Li:2017,Low:2016}, and quantum information processing technologies\cite{Tame:2013,Pelton:2015,Bozhevolnyi:2017b}. Another application recently gaining in popularity is the tuning of a surface's spectral reflectivity by plasmonic nanostructures to produce colour images\cite{Kristensen:2016} with ultra-dense information storage\cite{zijlstra:2009,Kumar:2012,Zhu:2016}. However, due to the nanometre size scales needed these structures often require elaborate fabrication methods such as electron-beam lithography (EBL) or focused ion beam (FIB) milling\cite{Fang:2013}. EBL allows for precise and reproducible definition of nanostructures but at the cost of time consuming process steps to produce a mask pattern in a polymer resist\cite{Chen:2015}. Additionally, in most state of the art lithography systems the spatial resolution is still limited to about 10\,nm\cite{Chen:2015}. FIB offers an alternative method for high-resolution and mask-less fabrication, but still requires long pattern writing times, and also comes with the potential problem of contaminating the structure materials with the milling ions\cite{Kim:2012}.

Self-assembled and self-similar metallic structures where high levels of field enhancement are hosted by the naturally occurring sub-nanometre sized gaps or protrusions, are thus of great interest as they offer an alternative source of strong field localization and enhancement. Typically these structures also have fast and scalable (bottom-up) fabrication methods with few processing steps\cite{Colson:2013,Klinkova:2014,Xi:2015}. While these types of structures promise scalable and easily fabricated nanostructures, their specific optical properties are however limited within the scope of their assembly methods. One way to expand the range of achievable excitations is of course to influence the assembly process during fabrication\cite{Gwo:2016}, but another way is to alter metallic and dielectric nanostructures post-assembly via controlled photothermal reshaping\cite{Shalaev:2007,Hu:2012,Garnett:2012,Herrmann:2014,Mertens:2016,Zhu:2016,Zhu:2017}. One such system, that we study here, is thin gold films near the percolation threshold subjected to morphological modifications by pulsed laser illumination. The nanostructured morphology of percolation metal films arises from the Volmer--Weber process of metal growth on dielectric substrates\cite{Greene:2010}. During the deposition process the metal atoms have a mutually strong interaction, while interacting less with the substrate. This leads to the formation of isolated clusters that tend to grow in the substrate plane during deposition, eventually reaching a percolation threshold where they merge to form a connected system. Further metal deposition will then serve to close up any remaining gaps in the film morphology, and eventually the system will transition into a metal on metal growth process. Depending on the deposition parameters and substrate used, it is possible to routinely fabricate large-scale areas of such metal structures where the smallest feature sizes can be at the sub-nanometre scale\cite{Greene:2010}. 

These films also hold interesting non-linear optical properties with prospects for white-light generation with modest pump powers\cite{Novikov:2017}. However, the optical properties of these films remain to be fully investigated experimentally with dedicated spectroscopic methods. The small feature sizes of these metal films present a significant challenge for experimental exploration with conventional near-field optics. For scattering scanning near-field optical microscopes (s-SNOMs) the spatial resolution limit of the setup will still be defined by the cantilever tip size, which in the most state of the art devices is still in the order of a few nanometres\cite{Wu:2006}. Despite being an exceptionally high spatial resolution for photon-based characterization, the limited spatial resolution of the tip does not allow to explore the optical properties of ultra-small plasmonic gaps with dimensions on the single nanometre scale and below\cite{Ducourtieux:2001,Genov:2003}. Furthermore, s-SNOM techniques are usually spectrally limited to a single or a small set of energies which is detrimental to fully characterize the plasmonic eigenmodes of these complex structures. Electron-based spectroscopic techniques such as energy-filtered electron energy-loss spectroscopy in monochromated transmission electron microscopes (TEMs) offers an attractive alternative to access the optical properties of nanometre and subnanometre sized plasmonic structures\cite{Scholl:2012,Scholl:2013,Raza:2014a,Raza:2015a,Raza:2016,Hobbs:2016,Bosman:2011,Losquin:2013,Gritti:2017}. Indeed, the exceptional subnamometre spatial resolution and the broadband excitation spectrum of the electron beam\cite{Abajo:2010,Colliex:2016} has recently been used to spatially map the intrinsic plasmonic eigenmodes in self-similar pristine silver films\cite{Losquin:2013}. This method permits to identify unambiguously plasmonic hot spots that are not related to the morphology of the films in a simple way, as predicted theoretically\cite{Seal:2003}.

Here we explore novel aspects of the optical reconfiguration of the plasmon excitations in gold films near the percolation threshold. After illumination with a fs-pulsed laser above a critical power, it is possible to induce permanent morphological modifications in the films. These changes allow for controlled changes to the film's optical properties. Using optical transmission spectroscopy, we demonstrate how it is possible to inscribe polarized and wavelength selective field enhancement into these otherwise broadly resonant films with laser writing. We explain the morphology changes of the films from laser illumination by photothermal reshaping of the metal nanoparticles in resonance with the excitation wavelength. We show that this process depends intricately on the degree of film percolation, illumination power, polarization, and wavelength of the laser. Our explanation includes three different processes: Particle spherification for elongated nanoparticles, dimer decoupling for large plasmonic gaps, and dimer welding/particle fusion for small gaps. We use hyperspectral imaging with nanometre resolution to show statistically the effect of morphological changes on the distribution of plasmonic modes in the near-field. By utilizing full-wave simulations of our explored morphologies, we highlight that the polarization dependence in the inscribed films originates from resonant elongated particles that have formed during the photothermal processes. Despite the complexity of the optical properties in these percolation films, we show that the far-field optical properties can be modified controllably through laser illumination. 

\section*{Results}
\subsection{Morphology changes and induced optical anisotropy.}
\begin{figure}[h]
\center
\includegraphics[width=16cm]{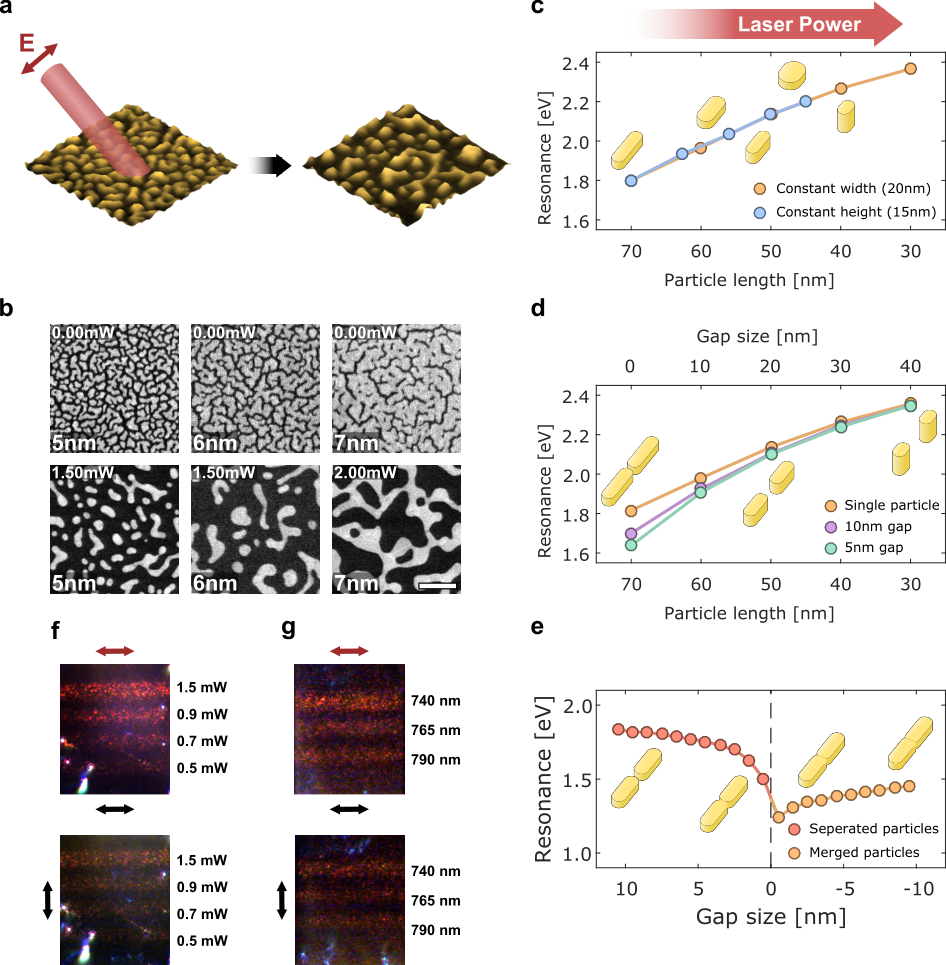}
\caption{\textbf{Optical reconfiguration of thin gold films.} (\textbf{a}) Schematic overview of the laser illumination process used to alter sample morphology. (\textbf{b}) STEM dark-field images of intrinsic film morphologies and laser illuminated morphologies. All images at same scale, scale bar is 150\,nm. (\textbf{c}) Simulations of the blueshift of longitudinal resonance energy for a particle being shortened, and then either widened or increased in height to preserve particle volume. (\textbf{d}) Simulation results of the blueshift from opening of a gap between two plasmonic particles from shortening the particles. (\textbf{e}) Simulation of how closing of a plasmonic gap and merging the two particles (welding) causes a redshift in resonance energy. (\textbf{f}--\textbf{g}) Transmission dark-field images of stripes written in the gold films with (\textbf{f}) different laser powers with a 740\,nm wavelength, and (\textbf{g}) with different laser wavelengths. Red arrows indicates polarization of laser writing, black arrows the polarization used when recording the image.}
\label{fig:1}
\end{figure}

We have fabricated 3 samples of thin gold films on 18\,nm thin SiO$_2$ TEM membranes (see methods). The films are of nominal thicknesses 5, 6, and 7\,nm. After fabrication, a series of areas on the gold films are illuminated by scanning a fs-pulsed laser at various powers across the films, Fig.~\ref{fig:1}.a (see methods). If the illumination is performed above a critical level of laser power, permanent morphological changes will be induced in the films. The degree and nature of this morphology reconfiguration can be controlled based on the scan parameters used, but principally depend on the laser power or wavelength used. Fig.~\ref{fig:1}.b shows scanning transmission electron microscope (STEM) dark-field images of the intrinsic film morphologies, and examples of the effect of laser illumination (a more detailed morphology study is available in supplementary Fig.~1). In Fig~\ref{fig:1}.b we observe that the reconfiguration of the gold films is remarkably different for percolated (7\,nm) and unpercolated (5\,nm) samples. Indeed, high powers in the writing laser will generate more isolated nanoparticles in gold films with thicknesses below the percolation threshold. This is in stark contrast to the percolated metal film where the film is already an inter-connected network of clusters, and it remains a network of connected gold clusters even for much larger laser powers.

To understand how the morphology changes are linked to changes in the plasmonic resonances in the intrinsic film, we present a simple toy model of resonant elongated particles that have their aspect ratios altered while maintaining their initial volume (see methods). We can imagine that three different scenarios will occur during the photothermal reshaping: particle shortening/spherification, particle decoupling/gap widening, and particle fusion/welding. In Fig.~\ref{fig:1}.c we highlight how shortening a particle while adding the 'lost' volume to either its height or width will result in a blueshift of the particle's longitudinal mode. This change in aspect ratios will correspond to the general contraction of a metal cluster into a more spherical shape due to the surface tension in the molten metal trying to minimize the particle's surface energy. In Fig.~\ref{fig:1}.d we show how this kind of particle contraction (lost volume in length added to height) of two coupled particles will again result in a blueshift of the system's longitudinal mode from increasing the gap distance between them. When the two particles are sufficiently contracted, their coupling will also be broken and they will start resonating like two individual particles. This kind of morphology change will correspond to how plasmonically coupled clusters that are too far apart for particle welding/fusion will become decoupled. Finally in Fig.~\ref{fig:1}.e we have the scenario of two metal clusters that are close enough to each other to fuse under the laser illumination. We here simulate the fusion process by progressively moving the two elongated particles closer to each other, and when they start to overlap (gap size of 0) the overlapping volume gets added to the resulting single particle's height. Both in terms of closing the gap distance and merging the two particles into one, a general redshift of the initial resonance is observed, with a strong shift at the merging point. After merging, the combined particle exhibits single particle behaviour (see Fig.~\ref{fig:1}.c).

The effect of the laser illumination on the gold films and its morphological changes can be observed with a microscope at low magnification, as shown in Fig.~\ref{fig:1}.f and g. From Fig.~\ref{fig:1}.f and g we see a clear difference between the illuminated and the unilluminated regions of the gold films from the distinctive colours that emerges from the morphological changes in the film. Interestingly, from Fig.~\ref{fig:1}.f and g, we also see that the laser writing has left behind a polarization dependence in the films' optical properties, correlated with the polarization of the laser used to perform the writing, as a redder hue becomes visible for the polarization aligned with the one used in the laser writing.

\subsection{Optical far-field properties of optical reconfigured gold films.}
To investigate the emergent red colour from laser illumination in greater detail, we perform optical spectroscopy on our samples with an inverted microscope connected to a state of the art spectrometer (see methods). From the bright-field transmission spectra of Fig.~\ref{fig:2} (full data available in supplementary Fig.~2), when aligning the polarization of the light source in the transmission experiment parallel with the polarization of the laser used for the writing, we see a sharp decrease in transmission for photon energies in the range of 1.8--2.0\,eV. When turning the polarizer to the perpendicular direction of the laser writing, this feature is strongly suppressed. On Fig.~\ref{fig:2}.c we show for the 5\,nm sample that the position of this feature is also dependent on the wavelength used for the illumination, with longer wavelengths also producing a feature deeper in the red part of the spectrum. For the 7\,nm (and 6\,nm) sample this feature is less pronounced for the lower powers used, but it still becomes apparent for sufficiently high powers. This can be understood from comparing the morphologies in Fig.~\ref{fig:1}.b, as the thicker gold films seem generally less perturbed by the laser illumination. We explain this from the fact that the 6 and 7\,nm films appear to be above the percolation threshold, and as such higher levels of laser power is needed to fully separate the particles as heat will be more efficiently conducted away from its injection point in a fully connected system\cite{Andersson:1976,Namba:1970}. Another observed trend is that the central position of the transmission feature tends to blueshift for increasing levels of laser power used for the writing. We attribute this to two effects. First, the individual particles become shortened for the higher powers of illumination, due to increasing spherification of the gold at higher temperatures, and secondly due to the likewise increasing thickness of the gold particles as the volume of gold that was previously covering the flat surface now accumulates into thicker particles (when we assume only minimal gold evaporation).

\subsection{Statistical analysis of EELS intensity distributions.}
We have shown that laser-induced reshaping of the metal films has a strong effect on their optical far-field properties. In order to investigate the plasmonic properties of the individual particle clusters and gaps, we recorded high resolution EELS maps of the samples. Due to the highly random nature of the film morphologies, the specific detailed distribution of the electric fields associated with the plasmon resonances is difficult to reproduce, making qualitative comparisons between samples difficult (example EELS maps are available in Fig.~3.a). Secondly, comparisons between large sets of EELS maps for such random geometries can become very overwhelming to interpret. But, due to the self-similar and isotropic nature of the films, a quantitative statistical analysis of sufficiently large regions of the films can provide a reproducible and comparable probability distribution function (PDF) of the EELS intensity of a specific film morphology. In order to construct PDFs of our EELS intensities, we take advantage of the methods of previous theoretical and experimental works on self-similar and fractal structures\cite{Stockman:1994,Bozhevolnyi:2001,Genov:2003}. In short, for each spectral image we can extract the resonance energy and EELS intensity of the measured plasmons. By then binning the found intensities we form a distribution of intensities in the spectral image (i.e. a histogram). For each intensity value, this PDF then gives us the probability of finding this EELS intensity in the image  (see methods for more details). The EELS intensity gives the probability of an electron losing a certain amount of energy along its trajectory through the sample, and it is proportional to the integral of the electric field induced in the sample along this trajectory by the electron\cite{Abajo:2010}. As such, the EELS intensity is not itself a direct measure of the plasmonic electric field amplitude, but the two are strongly related\cite{Abajo:2010,Kociak:2014,Shekhar:2017,losquin:2017}. We use this to justify the comparison of EELS intensity to previous measurements and calculations of the near-field intensities of semi-continuous metal films.

To extract the plasmon resonant energies and peak intensities from our background-corrected spectral images, we perform an iterative series of Gaussian fits on the individual spectra. First a set of energy ranges of interest are determined from taking an average of the full spectral image, and within these energy ranges individual Gaussian functions are fitted to a smoothing spline constructed from the data. The parameters of these fits are then used for initial guesses to fit a sum of Gaussian functions that are fitted to the full range of the individual spectra's datasets (see methods for more details). An example of these sequential fits can be seen in Fig.~\ref{fig:3}.b. From the extracted peak central energies we can construct histograms of how the resonances are concentrated in terms of energy, and from the found peak intensities we can construct distribution functions of the EELS intensities in terms of their spectral position (Fig.~\ref{fig:3}.c--f, and supplementary Figs.~3-5). For all of the films we see that prior to optical reconfiguration the central energies of the detected plasmon resonances form a wide continuum with a slight peak in the distribution around 2.1--2.2\,eV. The position of this peak in the distribution seems to redshift slightly for increasing film thickness. The constructed PDFs for the intrinsic films also show that the near-infrared and red part of the spectrum (1.3--1.8\,eV) contributes with higher relative EELS intensities, in agreement with prior experiments and the scaling theory for semi-continuous metal films predicting higher field enhancement for longer wavelengths\cite{Genov:2003,Sarychev:1999,Sarychev:2000}. 

As the samples gets subjected to high laser powers, we see a dramatic redistribution of the resonance energies, as well as the intensity distributions (results for the full range of powers on the three film samples are available in supplementary Figs.~3--5). The effect is especially pronounced for the 5\,nm sample, Fig.~\ref{fig:3}.c and e, where increasing levels of power of illumination gradually reshapes the PDFs from the normal distribution expected for these kinds of films, towards a scaling power-law of isolated dipoles\cite{Genov:2003,Stockman:1994}. We explain this from the morphology images in Fig.~\ref{fig:1}.b and our toy model in Fig.~\ref{fig:1}.c--e. For the high laser powers we see the inter-particle gaps increase significantly, and the resulting morphology consists mainly of isolated particles. As a result, the inter-particle coupling present in the intrinsic films has been lifted. This isolating and the reshaping into thicker and more spherical particles also explains why we generally see a strong blueshifting in the resonance energies in Fig.~\ref{fig:3}.c. For the 7\,nm sample we see a distinctly different morphology and resonance behaviour, compared to the 5\,nm sample. For all levels of laser power the 7\,nm sample remains a connected structure forming large networked clusters, where the small gaps that previously dominated the morphology either have fused or opened fully. The removal of these small features and their replacement by large connected clusters can help us to understand why the illuminated film seems to have a large increase in red and near-infrared modes in Fig.~\ref{fig:1}.d. From the fact that the structures also never become truly isolated, we can understand why the EELS intensity distributions for the illuminated parts do not deviate as dramatically in their shape from the intrinsic PDFs, as in the 5\,nm case. Interestingly, we are able to see a large increase in the relative EELS intensities from the $\sim$1.9\,eV part of the spectrum in the illuminated parts of the 5\,nm film, when compared with the intrinsic films. This increase in intensity also seems to become larger for increasing powers of laser illumination, and it is worth noting that the energy corresponds to the reduction in transmission observed for the parallel polarization for the 5\,nm sample in Fig.~\ref{fig:2}.a.

\subsection{Polarization dependence.}
In order to qualitatively investigate the polarization dependence of the extinction features seen in the transmission spectra of Fig.~\ref{fig:2}, we have constructed EELS maps for the 5\,nm film samples by integrating the EEL spectra in the 1.8--2.0\,eV energy range, as seen in Fig.~\ref{fig:5} (full energy ranges are available in supplementary Figs.~7--10). From these maps several elongated particles appear to host dipole-like plasmon resonances, that are predominantly aligned with the direction of the polarization of the laser used to induce the optical reconfiguration. To verify if these particles are dominated by a dipolar response in this energy range, we performed finite-element simulations of plane wave excitations with two orthogonal polarizations on the same film morphologies (see methods). A comparison between the simulated field distributions and the measured EELS intensities can be seen in Fig.~\ref{fig:7}. Because the plane wave simulations allow us to make comparisons to the two excitation polarizations individually, we can now see how the measured plasmon resonances from the EELS data are decoupled in terms of polarization. With this we show that the suspected elongated particles are in fact behaving like strongly polarized nanorods with resonances in the same energy range as the features measured in the transmission experiments. A one-to-one comparison of EELS and plane-waves is usually difficult due to the inherent differences of the excitation sources\cite{Husnik:2013}. The electron beam in EELS is able to excite plasmon modes that are normally inaccessible with optical fields\cite{Hohenester:2009}, and is also able to excite all different polarizations simultaneously\cite{Abajo:2010,Kociak:2014}. However, here we are only interested in identifying the bright modes around 1.9\,eV that are optically active in our transmission measurements. We expect from this comparison to identify the polarization dependence of the plasmonic modes in the reshaped films (Fig.~\ref{fig:2}). 

\section*{Discussion}
In summary, we have characterized the morphological and plasmonic restructuring of semi-con\-tinuous gold films of nominal thicknesses 5, 6, and 7\,nm when subjected to fs-laser pulses of different powers (Fig.~\ref{fig:1} and supplementary Fig.~1). This kind of optical reconfiguration becomes apparent in the far-field properties of the films, as a resonant feature is seen in the transmission spectra from the laser illuminated areas of the films. When the samples are illuminated with a polarized light source that is aligned to the polarization of the laser used for the optical reconfiguration, a strong decrease in transmission is observed. Furthermore, the spectral position of this transmission feature can be controlled by varying the laser wavelength and power when performing the optical reconfiguration (Fig.~\ref{fig:2}). By measuring EELS maps from selected regions of the altered and intrinsic parts of the samples, we have constructed histograms of the central energies of the plasmon resonances present in the films, as well as PDFs of the EELS intensity distributions for different resonance energies (Fig.~\ref{fig:3} and supplementary Figs.~2--4). From these we conclude that generally the resonances at longer wavelengths have higher EELS intensities in the intrinsic films, while in the altered films we observe a redistribution of the resonances that contribute to the largest EELS intensities. The intensities in reconfigured films are also generally higher overall, when compared to the intrinsic films. Finally, we have investigated the origin of the polarization dependence of the features observed in the transmission experiments, by plotting EELS intensity maps of energies related to the measured decrease in transmissions (Fig.~\ref{fig:5}). We have performed finite-element simulations to reconstruct the electric field distributions in our sample morphologies from polarized far-field excitation. We compare these to the measured EELS intensities to confirm that the resonant particles formed after laser illumination are responsible for the polarization dependence of the measured transmission spectra (Fig.~\ref{fig:7}).

These types of metallic films have previously been demonstrated to function as SERS sub\-strates\cite{Shalaev:2007,Gadenne:1997,Drachev:2005,Perumal:2014,Novikov:2016}. We show here that it is possible to convert the broad ensemble of different plasmonic resonances found in these films into a more narrow band of resonances, making the films more selectively resonant for a specific wavelength and polarization of light. This could potentially be used to further increase the enhancement factor from such films in SERS applications, by tuning and enhancing the resonances towards the wavelength and polarization of the laser used in the Raman experiment. New sensing applications could also harvest the potential of these kinds of metal films.

Finally, we have shown that by using a different laser wavelength or power when performing the optical reconfiguration it is possible to tune the wavelength of the resulting resonant particles created in the film. This could be useful for plasmonic colour printing by laser illumination, as it would be possible to inscribe pixels of different colours into the films by use of different wavelengths and powers, constructing a colour image that would be visible clearly when viewing the film through a correctly aligned polarizer. Experiments have already been performed demonstrating such concepts\cite{Zhu:2016}, but rely on using specifically fabricated substrate structures. The ease with which these metal films can be fabricated, even at very large-scale areas, could offer an alternative low-cost substrate structure for emerging plasmonic colour printing technologies, as well as applications in ultra-dense data storage media\cite{zijlstra:2009}. 

\section*{Methods}
\subsection{Fabrication.} Thin gold films of 5, 6, and 7\,nm nominal thicknesses were deposited onto 18\,nm thick SiO$_2$ TEM membranes from Ted Pella, Inc. using an electron beam deposition system. The gold was deposited with a constant rate of 2\,\AA/s, with the total deposition time defining the thickness of the final film. Chamber pressure was maintained at $\sim10^{-5}$\,mbar and deposition was on room temperature substrates.

\subsection{Optical reconfiguration.} The laser illumination is performed with a two-photon photoluminescence (TPL) setup, consisting of a scanning optical microscope in reflection geometry built based on a commercial microscope with a computer-controlled translation stage. The linearly polarized light beam from a mode-locked pulsed (pulse duration $\sim$200\,fs, repetition rate $\sim$80\,MHz) Ti-Sapphire laser (wavelength $\lambda$ = 730-–860\,nm, $\delta\lambda$ $\sim$10\,nm, average power $\sim$300\,mW) is used as an illumination source at the fundamental harmonic (FH) frequency. After passing an optical isolator (to suppress back-reflection), half-wave plate, polarizer, red colour filter and wavelength selective beam splitter, the laser beam is focused on the sample surface at normal incidence with a Mitutoyo infinity-corrected long working distance objective (100$\times$, NA = 0.70).  The FH resolution at full-width-half-maximum is $\sim$0.75\,$\upmu$m. The half-wave plate and polarizer allows accurate adjustment of the incident power. The laser illumination was done with the following scan parameters: Integration time (at one point) of 50\,ms, speed of scanning (between the measurement points) of 20\,$\upmu$m\,s$^{-1}$, and scanning step size of 350\,nm. The incident power was within the range 0.5--2.0\,mW. Unless stated otherwise, the excitation wavelength was fixed at 740\,nm. We should mention that with this laser power, morphology changes could only be induced in the pulsing regime of the laser.

\subsection{Optical spectroscopy.} Bright-field transmission spectra were recorded from the intrinsic and laser illuminated regions of the samples on the TEM membranes. A custom spectroscopy setup built from a Nikon Eclipse Ti-U Inverted microscope was used. The system is fitted with a halogen white-light source with peak emission at 675\,nm. The light is collected by a CFI S Plan Fluor ELWD objective from Nikon (60$\times$, NA = 0.70), and the spectra are recorded using a Shamrock 303i Spectrometer equipped with a Newton 970 EMCCD. A LPVISE200-A visible linear polarizer from Thorlabs is placed between the light source and sample stage, allowing for polarized illumination. Spectra are collected for polarizations perpendicular and parallel to the optically induced changes in the gold films. The transmission spectra are corrected for the spectral profile of the halogen lamp by recording a reference spectrum through a similar glass slide as the membranes are mounted on, for both polarizer positions. After dividing the intensities point by point with the reference, the final spectrum is obtained by averaging over the individual laser modified areas ($\sim$20 pixels on the CCD) and normalizing with the maximum value for comparison.

\subsection{STEM and EELS measurements.} The EELS measurements were performed with a FEI Titan TEM equipped with a monochromator and a probe aberration corrector. The microscope was operated in STEM mode at an acceleration voltage of 120\,kV, with a probe diameter of 0.5\,nm and with zero-loss peak (ZLP) full-width-half-maximums of $\sim$0.2\,eV. For spectral images, 500$\times$500\,nm$^2$ areas were scanned with 5--8\,nm step sizes, using larger step sizes for the films illuminated at greater laser powers. Before imaging, the samples were cleaned in an O$_2$-plasma for 45\,s.

From the width of the ZLP, its central position is determined and the energy scale of the spectra is shifted to zero for this position. The spectra are then normalized by integrating the background signal found in the higher energies of the spectrum. After normalization the ZLP is subtracted by performing a series of power law fits, and the fit producing the minimum residual is picked. The full background corrected spectral image is averaged together, and a sum of Gaussian functions are fitted to the now emerging 'primary modes' of the image, i.e. the modes so sufficiently common or intense that they contribute overwhelmingly to the total average of the spectral image. A smoothing spline is then constructed for each individual spectra's data. These smoothing splines are then segmented within the full-width-half-maximum of the Gaussian functions fitted to the 'primary modes', and within each of these data segments a Gaussian function is fitted. The quality of these individual fits are then evaluated to determine if a peak is present in this energy range of the spectrum or not. From each segment of the spectrum, we now get the information if a peak is present, and if so, what its amplitude and central position is. Using these values for initial guesses, a sum of Gaussian functions are then fitted to the full range of the spectrum's dataset and the plasmon resonance energies and EELS intensities are extracted from these fits.

\subsection{Statistical analysis.} Using the central position of the the Gaussian fits described above, a histogram can be constructed of identified central energies, with bin sizes of 0.01\,eV, matching the energy resolution of the TEM's detector. The PDFs of discretized windows of the energy range are constructed from the corresponding amplitudes of the Gaussian fits with central energies in those windows. Each energy window is 0.10\,eV wide, and to construct the PDF the fitted amplitudes are binned into 30 equally wide bins of EELS intensities. We have estimated a base level intensity as the median of the 50 lowest identified amplitudes from each of the spectral images for each sample. All the identified intensities for each energy window are then normalized with respect to this base level intensity to be represented in terms of a relative EELS intensity.

\subsection{Simulations.} 
Full-wave 3D simulations of the gold structures were carried out using the finite-element software package COMSOL Multiphysics (v5.1). The simulation domain was truncated with perfectly matched layers from all sides. The system was solved in a scattered wave formulation, with analytical solution for three layer system (air-glass-air) as the background field. A custom Python script was used, which generated the simulation geometry from the outlines that were extracted from the binaries of STEM dark-field images (see supplementary methods). The particles were represented as straight prisms in the 3D geometry. The relative thicknesses, $t_r$, of the gold films have been calculated by $t_r=\log\left(I_{tot}/I_{ZLP} \right)$, where $I_{ZLP}$ is the integral of the ZLP in the EEL spectrum, and $I_{tot}$ is the total integral of the EEL spectrum, ranging from 0--17\,eV. Calculating this for each pixel in our spectral images, and performing a plane background correction to this, provided the relative height map for the samples.
By averaging the height map values inside the particle contours, the relative heights of the particles were obtained. The tallest particles were set to a height of 28\,nm, as this provides the best correlation to the EELS measurements. An example of obtained particle contours and their relative heights are shown in supplementary Fig.~6.

Simulations for Fig.~\ref{fig:1}.c--e are based on a similar setup. One or two nanorods are placed on the glass membrane. Calculations for Fig.~\ref{fig:1}.c are done with a single nanorod that is composed of a rectangular prism (width $w$, height $h$, length $l$) with semi-cylindrical caps (radius $w/2$) at both ends. A series of calculations was done, for various lengths of the particle. For each particle size an absorption spectrum was calculated and the corresponding resonance peak was found. The particle size was varied to model the melting process, but the particle volume ($V=h\cdot\left(w l + 2 \pi w^2/4\right)$) was kept constant either by scaling the particle height ($h$) or particle width ($w$).

For Fig.~\ref{fig:1}.d we simulated two nanorods, with fixed centre positions. Consequently, as the particle length was reduced the gap between the nanorods increased. Particle height was scaled to keep the volume of the particles constant as they decreased in length.

For Fig.~\ref{fig:1}.e the particle length was kept constant but the distance between the two particles varied. For positive gap sizes the particle shape stayed constant, but for negative gap sizes (particle overlap) the particle height was increased to keep the volume of the merged particle equal to the volume of the two initial particles.

\newpage
\section*{References}
\bibliographystyle{naturemag}
\bibliography{references}

\begin{addendum}
\item[Acknowledgments] The authors gratefully acknowledge financial support from CONACyT Basic Scientific Research Grant 250719, and the Danish Council for Independent Research--Natural Sciences (Project 1323-00087). S.~I.~B. acknowledges the European Research Council, Grant 341054 (PLAQNAP). N.~A.~M. is a Villum Investigator supported by Villum Fonden, Grant 1401500. Center for Nano Optics was financially supported from the University of Southern Denmark (SDU 2020 funding), while Center for Nanostructured Graphene (CNG) was funded by the Danish National Research Foundation (CoE Project DNRF103). T. R. acknowledges support from Archimedes Foundation (Kristjan Jaak scholarship) and Villum Fonden (DarkSILD project).

C.~F. would like to personally thank S\o ren Raza, K\aa re Wedel Jacobsen, and Johan Rosenkrantz Maack for many valuable discussions during this work.

\item[Author contributions] N.~S. and C.~F. performed the fabrication and EELS measurements, and C.~F. performed the EELS and image data analysis. T.~R. performed the numerical simulations. M.~G. performed the transmission measurements and data analysis. S.~N. and J.~B. performed the laser illumination of samples. A.~L., S.~X., S.~I.~B., N.~A.~M., and N.~S. supervised the project. C.~F. drafted the manuscript, and all authors contributed to its writing.

\item[Conflict of interests] The authors declare no competing financial interests.

\item[Correspondence]  Correspondence and requests for materials should be addressed to C.~F. (email:\\chrfr@fotonik.dtu.dk) or N.~S. (email: niste@fotonik.dtu.dk).

\end{addendum}

\newpage
\linespread{1}
\begin{figure}[h]
\center
\includegraphics[width=14cm]{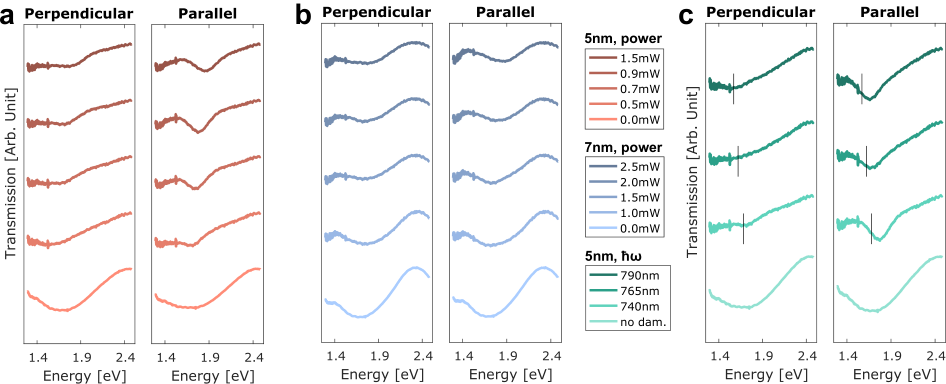}
\caption{\textbf{Optical anisotropy.} Polarized bright-field transmission spectra of the (\textbf{a}) 5\,nm, (\textbf{b}) 7\,nm films from regions illuminated with different laser powers, and (\textbf{c}) for illumation with different photon energies on the 5\,nm sample. Vertical black lines indicate the photon energy of the laser used. The white-light source used to illuminate the samples in the measurements was polarized parallel or perpendicular to the polarization of the laser used for reconfiguration.}
\label{fig:2}
\end{figure}
\begin{figure}[h]
\center
\includegraphics[width=16cm]{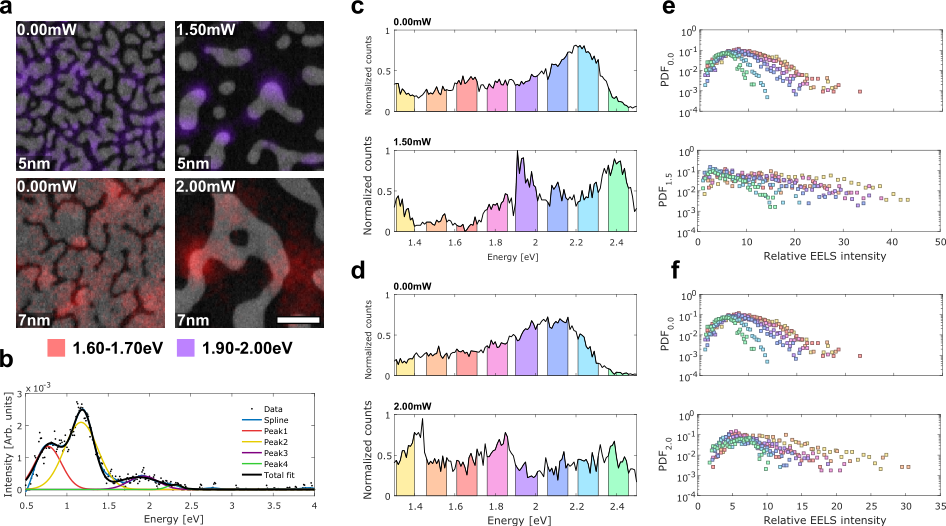}
\caption{\textbf{Plasmonic resonance distribution.} (\textbf{a}) Integrated EEL spectra overlaid on STEM dark-field images for 5 and 7\,nm morphologies. Scale bar is 75\,nm. (\textbf{b}) Example of an EEL spectrum and Gaussian fits used to extract peak central positions and energies. Normalized histograms of identified central energies of plasmon resonances in the (\textbf{c}) 5\,nm and (\textbf{d}) 7\,nm films before and after illumination. Constructed PDFs of the EELS intensity distributions for the intrinsic and reconfigured (\textbf{e}) 5\,nm and (\textbf{f}) 7\,nm samples.} 
\label{fig:3}
\end{figure}
\begin{figure}[h]
\center
\includegraphics[width=12cm]{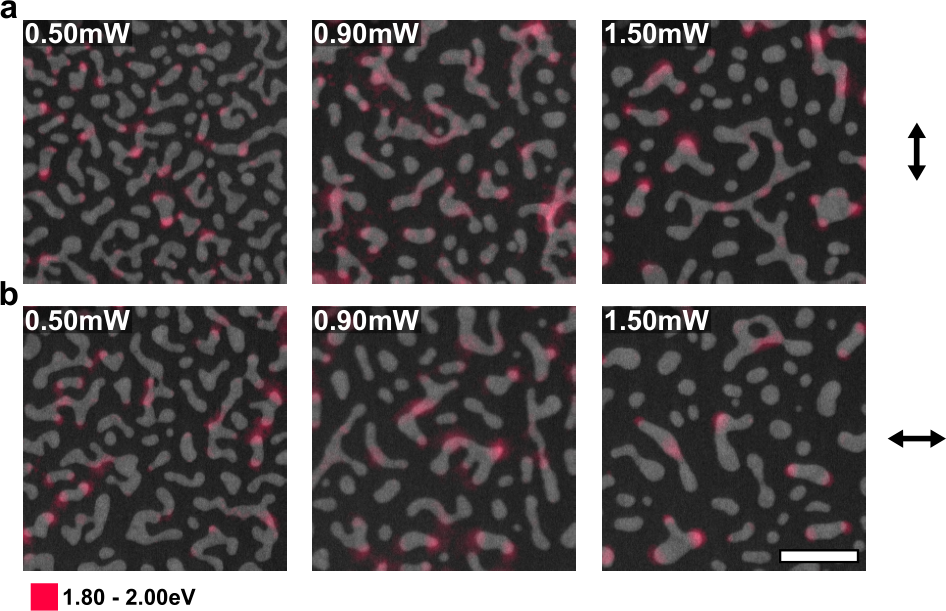}
\caption{\textbf{EELS anisotropy.} Colour maps of the recorded EEL spectra integrated in the energy range of 1.80--2.00\,eV and overlaid on top of their respective STEM dark-field images of the 5\,nm samples. Each row of images consists of samples with the same polarization used for illumination, (\textbf{a}) being the samples with $y$-polarized laser illumination, and (\textbf{b}) the samples with $x$-polarized illumination. The numbers in the top left corners of the images indicate the amount of power used for illumination. The scale bar is 150\,nm.}
\label{fig:5}
\end{figure}
\begin{figure}[h]
\center
\includegraphics[width=15cm]{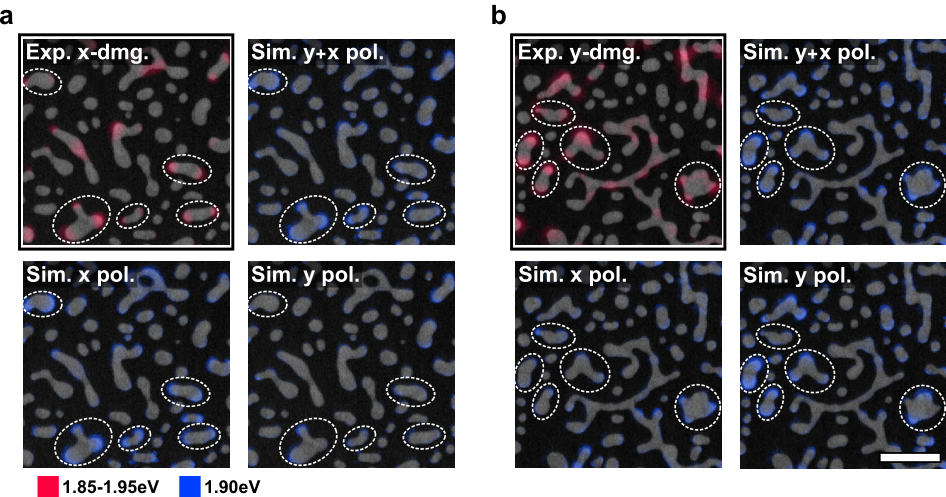}
\caption{\textbf{Far-field simulations.} Comparison between EELS data and the norms of electric field $z$-components from simulated polarized far-field excitation of the 1.5\,mW illuminated regions in the 5\,nm sample. (\textbf{a}) for the damage illuminated with polarization along the images' $x$-axis, and (\textbf{b}) for the sample illuminated along the $y$-axis. The EELS data was integrated in the region of 1.85--1.95\,eV and the excitation energy for simulations was 1.90\,eV. The electric field norms from the simulation were taken right below the gold particles. The scale bar is 150\,nm.}
\label{fig:7}
\end{figure}

\end{document}